# Carnahan Starling type equations of state for stable hard disk and hard sphere fluids

Hongqin Liu*

Integrated High Performance Computing Branch, Shared Services Canada, Montreal, QC, Canada

## Abstract

The well-known Carnahan-Starling (CS) equation of state (EoS)[1] for the hard sphere (HS) fluid was derived from a quadratic relation between the integer portions of the virial coefficients, $B_{n,integer}$, and their orders, $n$. Here we extend the method to the full virial coefficients $B_n$ for the general D-dimensional case. We assume a polynomial function of (D-1)$^{th}$ order for the virial coefficients starting from $n = 4$ and EoS's are derived from it. For the hard rob (D=1) case, the exact solution is obtained. For the stable hard disk fluid (D=2), the most recent virial coefficients up to the 10$^{th}$ [2] and accurate compressibility data[3,4] are employed to construct and test the EoS. For the stable hard sphere (D=3) fluid, a new CS-type EoS is constructed and tested with the most recent virial coefficients[5,2] up to the 11$^{th}$ and with the highly-accurate simulation data for compressibility[6-8]. The simple new EoS's turn out to be as accurate as the highest-level Padé approximations based on all available virial coefficients, and significantly improve the CS-type EoS in the hard sphere case. We also shown that as long as the virial coefficients obey a polynomial function any EoS derived from it will diverge at the non-physical packing fraction, $\eta = 1$.

* Email: hongqin.liu@canada.ca; hqliu2000@gmail.com

# I Introduction

The properties of high-density condensed phases are dominated by repulsive forces associated with the hard cores of constituents[9-12]. As a result, hard sphere systems can serve as model systems in various applications or references in the perturbation theories [10-13] for more realistic systems. Highly accurate descriptions of hard sphere systems are therefore of fundamental importance. In particular, equations of State (EoS) play a central role in calculations of thermodynamic properties and studies in this area have been very active for over one and half centuries[10-14]. Baus and Colot (1987) proposed a unified EoS for D-dimensional hard sphere fluids[9] and their work had considerable impacts on theoretical aspects of EoS developments and liquid structure descriptions. However, due to limited virial coefficients available and lack of highly-accurate compressibility simulation data at the time, the accuracies of the most EoS's constructed there are not high enough compared with those developed at later times[12,14].

Based on a generalization of the virial coefficient relation proposed by Carnahan Starling (CS) for their well-known EoS of hard sphere fluid[1], in this work, we propose a unified approach to the development of equations of state for D-dimensional hard sphere fluids. The focuses are the stable hard disk (HD, D=2) and hard sphere (HS, D=3) fluids. The goal is to derive simple CS-type EoS's with the "highest-possible" accuracy for the stable fluid region. Here the "highest-possible" accuracy refers to reproducing the compressibility data with accuracies comparable to that of the highest-level Padé approximation.

# II A generic approach to equations of state of D dimension hard sphere fluids

A generic equation of state for D-dimensional stable fluids was proposed by Baus and Colt (1987)[9]:

$$Z_D = \frac{1 + \sum_{n=1}^{\infty} c_n \eta^n}{(1-\eta)^D} \quad (1)$$

where $D$ is the dimension, $Z$, the compressibility factor ($Z = PV/(Nk_BT) = P/(\rho k_B T)$), $P$, the pressure, $T$, the temperature, $V$, the volume, $N$, the number of particles, $\rho$, the density, $k_B$, the Boltzmann constant, and the constants, $c_n$, are related to virial coefficients[9]. The packing fraction is defined as:

$$\eta = \frac{\pi^{\frac{D}{2}}}{\Gamma\left(1+\frac{D}{2}\right)} \left(\frac{\sigma}{2}\right)^D \rho \quad (2)$$

where $\sigma$ is the particle diameter, $\Gamma$, the gamma function. Eq.(1) is a scaled form of the virial series and the term $(1-\eta)^{-D}$ was empirically introduced to accelerate the convergence of the series. When the virial term in the numerator of Eq.(1) is truncated, somewhat arbitrarily, a CS-type EoS can be obtained. Here the CS-type EoS is defined as $Z_D = f(\eta)/(1-\eta)^D$, where $f(\eta)$ is a polynomial function. We start from the original virial EoS:

$$Z_D = 1 + B_2\eta + B_3\eta^2 + \sum_{n=3}^{\infty} B_{n+1}\eta^n \quad (3)$$

Apparently, in Eq.(3) the virial coefficients are dimension-dependent as well. In the following, for the HS case, the subscript "D" is dropped for convenience. The second and third virial coefficients in Eq.(3) are deliberately separated from the rest for the following constructions. If some simple correlation between $B_n$ and $n$ can be established, then a compact EoS will be derived from Eq.(3). A well-known example is the Carnahan-Starling (CS) EoS[1,10]. When only the integer portions of the virial coefficients are taken into account, a simple quadratic relation can be found:

$$B_n \approx B_{n,integer} = n^2 + n - 2, \quad (n = 2, 3, \ldots) \quad (4)$$

Combining Eq.(3) and (4) leads to the CS EoS[1]:

$$Z = \frac{1 + \eta + \eta^2 - \eta^3}{(1-\eta)^3} \quad (5)$$

This remarkably simple form enjoys acceptable accuracy, which makes the EoS the most adopted one for the stable HS fluid[10-14]. After careful analysis on the values of virial coefficients for the HD[2] and HS[2,5] fluids, it is found that a polynomial function of the order of $(D-1)$ can be used to correlate the virial coefficients:

$$B_n = \sum_{k=0}^{D-1} c_k n^k, \quad (n \geq 4) \quad (6)$$

Here we leave $B_2$ and $B_3$ to be incorporated in the final EoS and therefore excluded from eq.(6) to ensure the high accuracy in low density range. The coefficients, $c_k$, in Eq.(6) are different from those in Eq.(1). Now Eq.(3) can be written as:

$$Z_D = 1 + B_2\eta + B_3\eta^2 + \sum_{n=3}^{\infty} \sum_{k=0}^{D-1} c_k (n+1)^k \eta^n \quad (7)$$

For the hard rod case, D=1, $B_n = 1$, Eq.(7) reduces to the exact solution[9]:

$$Z_{D=1} = \frac{1}{1-\eta} = \sum_{i=0}^{\infty} \eta^i \quad (8)$$

Eq.(8) also servers as the base for deriving the final EoS's for other cases. For the hard disc case, ($B_2$=2), $B_n = c_1 n + c_0$, from Eq.(7) and (8) (taking derivative on Eq.(8)), after some straightforward manipulations, we have:

$$Z_{D=2} = \frac{1 + (B_3 - 3)\eta^2 + a_0\eta^3 + a_1\eta^4}{(1-\eta)^2} \quad (9)$$

The coefficients $a_0$ and $a_1$ are related to the coefficients, $c_1$ and $c_0$:

$$a_1 = 2(1 - B_3) + 4c_1 + c_0$$
$$a_2 = B_3 - 3c_1 - c_0 \quad (10)$$

For the hard sphere case, $B_n = c_2 n^2 + c_1 n + c_0$, ($B_2 = 4, B_3 = 10$). Eq.(7) and (8) (taking derivatives on Eq.(8)) lead to:

$$Z_{D=3} = \frac{1 + \eta + \eta^2 + a_0\eta^3 + a_1\eta^4 + a_2\eta^5}{(1-\eta)^3} \quad (11)$$

where the relationships between the coefficients are given by:

$$a_1 = -19 + 16c_2 + 4c_1 + c_0$$
$$a_2 = 26 - 23c_2 - 7c_1 - 2c_0$$
$$a_3 = -10 + 9c_2 + 3c_1 + c_0 \quad (12)$$

Similarly, we can derive an EoS for D=4, but we will not discuss the case here. To summarize, we have the following "generic" EoS for D-dimensional hard sphere fluids:

$$Z_D = \frac{1 + (B_2 - D)\eta + (B_3 - DB_2 + \sum_{i=0}^{D-1} i)\eta^2 + \sum_{i=0}^{D-1} a_i \eta^{i+3}}{(1-\eta)^D} \quad (13)$$

In the following, we will discuss the HD and HS cases in details, respectively. Greater attention will be devoted to the HS case. In a separate publication[15], Eq.(9) will be employed in phase transition analysis for the HD systems.

## III The virials coefficients and EoS for the stable hard disc fluid

For the stable HD fluid, Clisby and McCoy (2006)[2] published their calculated values for 9$^{th}$ and 10$^{th}$ virial coefficients and thus all the virial coefficients up to 10$^{th}$ are available. For the compressibility (pressure), computer simulation data have been reported by Kolafa and Rottner (2006)[3] and Erpenbeck and Luban (1985)[4]. These two sets of data (virial coefficients and compressibility) will be used for constructing and testing our new EoS. Namely, the coefficients $a_0$ and $a_1$ of Eq.(9) are determined by simultaneously fitting Eq.(9) with the two sets of data. For the HD fluid, the third virial coefficient $B_3 = 3.128$, $B_3 - 3 \approx 1/8$. The final EoS for stable HD fluid of this work reads:

$$Z_{D=2} = \frac{1 + \frac{1}{8}\eta^2 + \frac{1}{18}\eta^3 - \frac{4}{21}\eta^4}{(1-\eta)^3} \quad (14)$$

The conjugate virial coefficient correlation is given as:

$$B_n = 0.9931n + 0.3392, \quad (n \geq 4) \quad (15)$$

For the stable HD fluid, few dozens of EoS's have been published[12,16-18]. For a meaningful comparison, here we only consider few CS-type EoS's. For more detailed discussions, the reader is referred to Ref [12] and[17]. The simplest EoS is the Henderson EoS[16]:

$$Z_{D=2} = \frac{1 + \frac{1}{8}\eta^2}{(1-\eta)^2} \quad (16)$$

Eq.(16) is a special case of Eq.(14) by omitting the high order terms. Solana[17] proposed another simple EoS by introducing a new term in Eq.(16):

$$Z_{D=2} = \frac{1 + \frac{1}{8}\eta^2 - \frac{1}{10}\eta^3}{(1-\eta)^2} \quad (17)$$

More recently, Wang (2010)[18] proposed an empirical EoS for the HD fluid:

$$Z_{D=2} = \frac{1 + 0.128\eta^2 - 0.06\eta^4 - 0.11\eta^6}{(1-\eta)^2} \quad (18)$$

For a given set of virial coefficients, the Padé approximation is usually adopted to derive an equation of state. Since the Padé EoS reproduces all the virial coefficients exactly, it can be used to generate highly accurate compressibility (pressure) with the given virial coefficients. The general form of the Padé approximation reads[2]:

$$Z(M/L) = \frac{1 + a_1\eta + a_2\eta^2 + \cdots + a_M\eta^M}{1 + b_1\eta + b_2\eta^2 + \cdots + b_L\eta^L} \quad (19)$$

where $M + L = m$ and $m$ is the total number of the virial coefficients available. The constants, $a_1, \ldots, b_1, \ldots,$ are determined from exactly matching the virial coefficients. For the stable HD fluid, there are totally 9 virial coefficients reported[2]. Therefore, two equations, $Z(5/4)$ and $Z(4/5)$, can be derived. Using the virial coefficients from Ref [2], Wang (2010) obtained the coefficients for both $Z(5/4)$ and $Z(4/5)$[18]. By using the compressibility data from Refs [3,4], it is found that $Z(4/5)$ is slightly better than $Z(5/4)$. Therefore we will use $Z(4/5)$ for our comparison. The coefficients of the equation can be found in Ref [18] (Eq.(8) in the reference).

Figure 1 illustrates the virial coefficients from Ref[2] (up to the 10$^{th}$) and the calculation results from Eq.(15). The average absolute deviation (AAD) of the prediction from Eq.(15) for 4$^{th}$ to 10$^{th}$ coefficients is 0.67%. For verifying the linearity of Eq.(15), extrapolation up to $B_{16}$ is also depicted. The estimated extrapolation points are from Ref.[19]. The results show that the linearity holds up to high orders (>16$^{th}$).

**Table 1** a comparison of accuracies of EoS's

| EoS | Eq.(14) | Eq.(16) | Eq.(17) | Eq.(18) | Z(4/5) |
|---|---|---|---|---|---|
| AAD% | 0.16 | 0.78 | 0.23 | 0.16 | 0.15 |

The average absolute deviation: $AAD\% = \frac{100}{N_P}\sum_{i=1}^{N}\frac{Z^{cal}-Z^{sim}}{Z^{sim}}$.

Table 1 lists a comparison of compressibility predictions with the EoS's mentioned above. The data sources are Refs. [3] and [4] and density range tested is $0 < \rho^* \leq 0.86$. It can be seen that the Henderson EoS, Eq.(16), behaves worst among all the equations. Eq.(17) is slightly worse than the other three equations. Eq.(14) and (18) are as accurate as the Padé approximation, $Z(4/5)$. This means that with the given virial coefficients (up to 10$^{th}$), these two CS-type EoS's have reached the highest accuracy.

Figure 2 depicts the relative deviations ($RD\%$) over the entire density range considered. Firstly, we can see that the two sets of compressibility data are consistent with each other. Another immediate observation is that Eq.(14), (17), (18) and (19) all behave poorly as $\rho^* > 0.8$. In Ref[17], Mulero et al. (2009) introduced an extra term, $-50000\eta^{40}/(1-\eta)^3$, into Eq.(17) and the final EoS works very well up to $\rho^* = 0.89$. It has been reported that at density $\rho^* \sim 0.891$ there exists a liquid-hexatic phase transition and the compressibility exhibits a maximum at the point. The simple monotonic CS type EoS, Eq.(14) etc., cannot describe this special feature. A full account of the phase transitions in the HD system is addressed elsewhere[15].

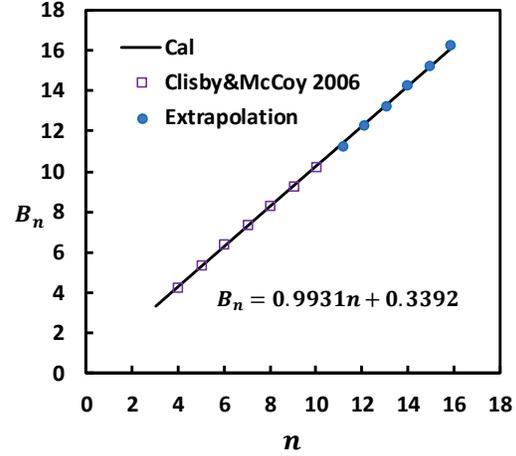

**Figure 1.** The virial coefficients of the hard disc fluid. Exact values are from (Clisby&McCoy 2006)[2]. Extrapolation points are from ref [19].

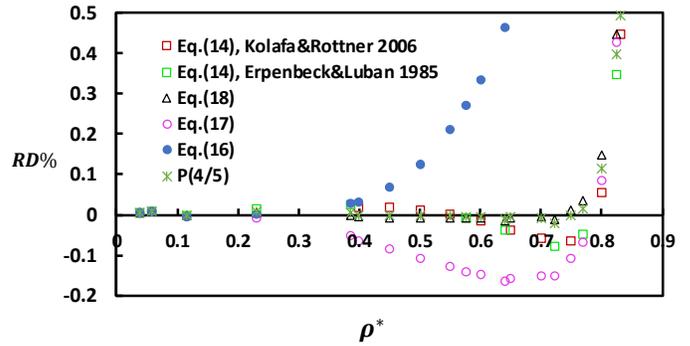

**Figure 2.** Plot of relative deviation vs reduced density for different equations, $RD\% = 100(Z^{cal} - Z^{sim})/Z^{sim}$.

## IV The virials coefficients and EoS for the stable hard sphere fluid

The HS fluid is one of the most important model fluids for both theoretical studies and practical applications[9-14]. Several dozens of EoS's have been proposed for the HS systems, including the stable fluid and metastable phase[6-14,20-24]. Although some EoS's have reached such high accuracies that the simulation data (compressibility) can be reproduced up to 6+ digits[6,7], their lengthy expressions limit the applications mainly to theoretical analysis. For most practical applications, compact EoS's or CS-type EoS's[1,9,10,12,22], are mostly employed. The simple formats of the CS-type EoS's, such as CS EoS (1969)[1], Kolafa EoS (1986)[2] and Baus-Colot EoS (1987)[9] allow one to easily derive other thermodynamic properties. Recent simulation data[6,7] show that there is still plenty of room for improving the accuracies of the CS-type EoS's.

In last decade or so, there have been significant progresses in the study of HS systems in several areas. Highly accurate virial coefficients up to 11th have been published (Glisby and McCoy, 2006[2]; Schultz and Kofke, 2014 [5]). For the pressure or compressibility, highly-accurate and consistent computer simulation data have been reported for both stable and metastable regions[6-8]. Very accurate EoS's based on truncated virial series have also been proposed[6,7,12,23,24]. The progresses in virial coefficient calculations and EoS data simulations make it possible for one to build a new CS-type EoS with high accuracies.

The most successful CS-type EoS (in terms of simplicity and accuracy) is perhaps the Kolafa EoS (1986), which was reported by Boublik and Nezbeda (1986)[10]. The CS correlation of the virial coefficients, Eq.(4), was slightly modified as:

$$B_n = \frac{5}{6}(n^2 + n) - 2, \qquad n \geq 3 \quad (20)$$

The derived EoS reads[10]:

$$Z = \frac{1 + \eta + \eta^2 - \frac{2}{3}(\eta^3 + \eta^4)}{(1-\eta)^3} \quad (21)$$

Eq.(21) can be considered as a special case of Baus and Colot's EoS (Eq.(2.22) in Ref [9]). Eq.(21) improves the accuracy in compressibility prediction by almost one order of magnitude compared to the CS EoS, Eq.(5). In their review paper [10], Boublik and Nezbeda (1986) also proposed an empirical CS-type EoS, which can be seen as a special case of Eq.(11). Since the coefficients of the EoS were obtained from limited virial coefficient and compressibility data, it turns out to be not better than the Kolafa EoS, Eq.(21). Consequently, it will be excluded from the following discussions. Other CS-type EoS's[9,12] are mostly not more accurate than Eq.(21) or with more complex forms and therefore will not be discussed either.

It has been a challenge for a simple compact EoS to have high precisions for both compressibility and virial coefficients. Only until recently, by building up from the isothermal compressibility route and using the most recent virial coefficients reported by Schultz and Kofke (2014)[5], Hansen-Goos (2016)[22] proposed a remarkably high precision EoS for the stable HS fluid:

$$Z = \frac{a \ln(1-\eta)}{\eta} + \frac{\sum_{i=0}^{7} b_i \eta^i}{(1-\eta)^3} \quad (22)$$

The parameters can be found in the reference (Table 1 in Ref [22]). Eq.(22) is probably the most accurate compact EoS (none CS-type) up to date for both compressibility and virial coefficients. Unfortunately, the logarithm term in Eq.(22) makes it inconvenient for some thermodynamic property calculations, such as entropy and chemical potential where integration is required.

For testing the compressibility predictions by using all available virial coefficients (up to the 11th, Ref[2,5]), the Padé approximation, Eq.(19), needs to be employed and 10 constants can be obtained. After some testing with the virial coefficients from Ref[5], it is found that $Z(5/5)$ provides best prediction for compressibility. Therefore, we will use $Z(5/5)$ in our discussions.

Moreover, after some comparisons, it is found that the recent virial coefficients from Schultz and Kofke (2014)[5] produce slightly better compressibility predictions. Therefore, we use this data set for determining the constants in Eq.(11). The other data set 2 is used for testing purpose.

Numerous computer simulation (MD, or MC) data have been reported for compressibility of hard sphere fluids[6-8,12,14]. After some analysis, it is found that the data sets from Pieprzyk et al. (2019)[6], Bannerman et al. (2010)[7] and Kolafa et al. (2004)[8] are most accurate and consistent with each other. As a result, the data set from Pieprzyk et al. (2019)[6] is employed for fitting the constants in Eq.(11) and the data sets from Ref[7,8] are used for testing. In summary, the constants in Eq.(11) are fitted simultaneously with the virial coefficients (up to 11th)[5] and the compressibility data[6], and the final EoS reads:

$$Z = \frac{1 + \eta + \eta^2 - \frac{8}{13}\eta^3 - \eta^4 + \frac{1}{2}\eta^5}{(1-\eta)^3} \quad (23)$$

and the conjugate virial coefficient relation follows:

$$B_n = 0.9423 n^2 + 1.28846 n - 1.84615, (n \geq 4) \quad (24)$$

Table 2 summarizes all the data sources for the fitting and testing. The results for virial coefficients are illustrated in Figure 3 and Table 3. Table 4 lists the coefficients of the Padé approximation $Z(5/5)$ calculated in this work from virial coefficients of Ref[5].

**Table 2** data sources for the HS fluids

| virial coefficients | Authors | highest order | |
|---|---|---|---|
| Fitting (SK) | Schultz&Kofke 2014 [5] | 11th | |
| Testing (CM) | Clisby&McCoy 2006 [4] | 11th | |
| **Compressibility** | Authors | density | $N_P$ |
| Fitting (set 1) | Pieprzyk et al. 2019 [6] | 0.05 - 0.938 | 59 |
| Testing (set 2) | Kolafa et al. 2004 [8] | 0.2 - 0.96 | 24 |
| Testing (set 3) | Bannerman et al. 2010 [7] | 0.1 - 0.96 | 35 |

$N_P$: number of data points.

**Table 3** Hard sphere virial coefficients, $B_n$

| $n$ | Schultz & Kofke 2014 | Eq.(25) | Clisby & McCoy 2006 |
|---|---|---|---|
| 2 | 4 | 4 | 4 |
| 3 | 10 | 10 | 10 |
| 4 | 18.36481 | 18.38462 | 18.3648 |
| 5 | 28.2244 | 28.15385 | 28.2245 |
| 6 | 39.8152 | 39.80769 | 39.8151 |
| 7 | 53.3421 | 53.34615 | 53.3444 |
| 8 | 68.5285 | 68.76923 | 68.5375 |
| 9 | 85.8259 | 86.07692 | 85.8128 |
| 10 | 105.682 | 105.2692 | 105.7751 |
| 11 | 126.488 | 126.3462 | 127.9263 |

AAD=0.19% vs SK[5] data, 0.34% vs CM[4] data.

**Table 4** coefficients of $P(5/5)$

| $b_1$ | -0.626902 | $a_1$ | 3.373098 |
|---|---|---|---|
| $b_2$ | 0.3514264 | $a_2$ | 7.843818 |
| $b_3$ | -3.842403 | $a_3$ | 9.659049 |
| $b_4$ | 3.208818 | $a_4$ | 8.064957 |
| $b_5$ | 0.0439983 | $a_5$ | 3.030368 |

Table 5 lists the comparison of AADs between the EoS's discussed here. The Padé approximations, $P(5/5)$, from two sets of virial coefficients, Ref [2] and Ref [5], respectively, show that the most recent data [5] produce slightly better prediction for the compressibility. The simple new EoS, Eq.(23), is as accurate as Eq.(22). In compressibility predictions, both Eq.(22) and (23) have same accuracy as the highest-level Padé approximation while sacrificing some accuracies in virial coefficient predictions.

For the three CS-type EoS's, when compared with compressibility data set 1 [6], the Kolafa EoS, Eq.(21) improves the prediction over the original CS EoS, Eq.(5) by one-order of magnitude; the new EoS, Eq.(23), over the Kolafa EoS, Eq.(21), by almost another order of magnitude. The new EoS also significantly improves the virial coefficient predictions compared with Eq.(5) and (21).

**Table 5** AAD% comparison between compact EoS's

| EoS | $Z$ | | $B_n$ | |
|---|---|---|---|---|
| | Set 1 | set 2&3 | SK | CM |
| CS EoS, Eq.(5) | **0.152** | 0.18 | 1.71 | 1.61 |
| Kolafa eq.(21) | **0.020** | 0.024 | 0.88 | 1.02 |
| Hansen-Goos eq.(22) | 0.0039 | 0.0074 | 0.18 | 0.31 |
| P(5/5), SK, Eq.(19) | 0.0042 | 0.0070 | 0.0 | 0.0 |
| P(5/5), CM, Eq.(19) | 0.0047 | 0.0076 | 0.0 | 0.0 |
| New EoS, Eq.(23) | **0.0037** | 0.0078 | 0.19 | 0.34 |

Figure 4 depicts the relative deviations from different EoS's over the density range up to the freezing point. The results from P(5/5), Eq.(22) and (23) overlap each other as expected. The prediction from a truncated virial series (up to 12$^{th}$, where $B_{12}$ is estimated from Eq.(25)) is also shown and it behaves poorly as $\rho^* > 0.6$.

Compared with Eq.(22) and $P(5/5)$, Eq.(23) is simpler and can be easily applied to calculations of various thermodynamic properties. For example, isothermal compressibility, $k_T$, can be calculated by the following equation:

$$k_T^* \equiv k_T\left(\frac{k_B T}{\sigma^3}\right) = \left(\eta \frac{\partial Z}{\partial \eta} + Z\right)^{-1} \rho^{*-1} \quad (25)$$

Where $k_T^*$ is a dimensionless reduced form, $\rho^*$, the reduced density, $\rho^* = \rho\sigma^3$ and

$$\frac{dZ}{d\eta} = \frac{4 + 4\eta - \frac{11}{13}\eta^2 - \frac{52}{13}\eta^3 + \frac{7}{2}\eta^4 - \eta^5}{(1-\eta)^4} \quad (26)$$

Figure 5 depicts the calculation results compared with the recent MC simulation data (Grigoriev et al., 2020)[25]. Chemical potential can also be easily calculated as well:

$$\mu^* \equiv \frac{\mu}{Nk_B T} = Z - 1 + I \quad (27)$$

where

$$I = \int_0^\eta \frac{Z_v - 1}{\eta} d\eta = \frac{188\eta - 126\eta^2 - 13\eta^4}{52(1-\eta)^2}$$
$$- \frac{5}{13}\ln(1-\eta) \quad (28)$$

By the way, the excess entropy reads:

$$s^{ex} = \frac{S - S^{id}}{Nk_B} = -I \quad (29)$$

Figure 6 depicts the chemical potential prediction compared with the simulation data. As expected, the predicted values are highly consistent with those from computer simulations[26-30].

Figure 3. Plot of the HS virial coefficients. The solid line is calculated by Eq.(24) by fitting the virial coefficients up to 11th (Schultz and Kofke, 2014 [5]). The results of Clisby and McCoy (2006) [2] are used for testing. Extrapolation points are from Hu&Yu (2009)[19].

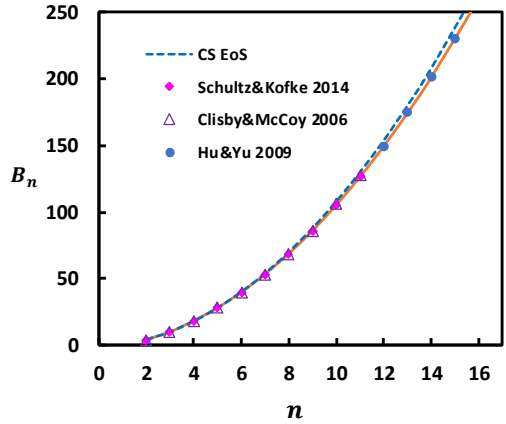

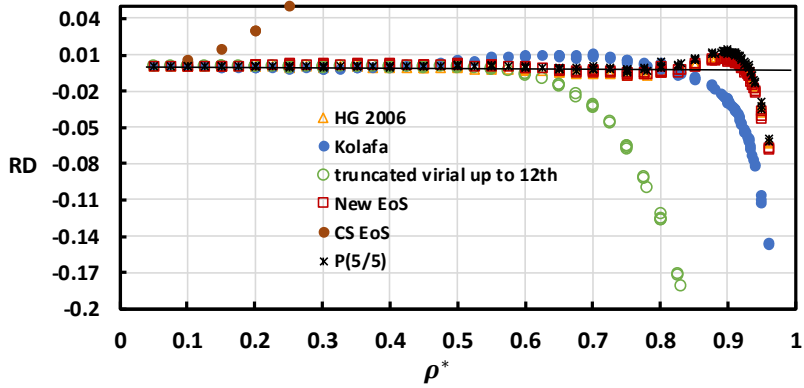

Figure 4. Comparison of relative deviations from different EoS's. The deviations from the CS EoS run off the grid after 6th virial coefficient. The results from P(5/5), (22) and (23) are almost the same. The 12th coefficient is estimated from Eq.(24), $B_{12} = 149.3$. For comparison, Padé extrapolation gets $B_{12} = 149.7$.

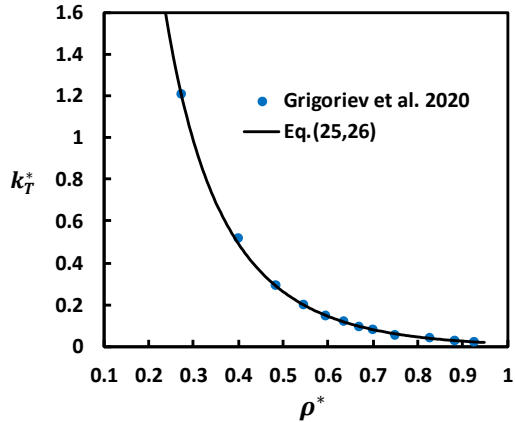

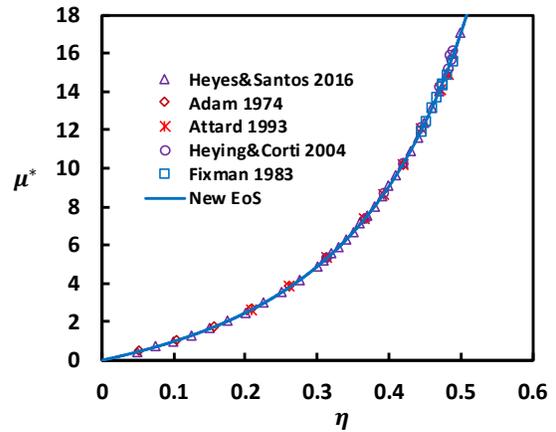

Figure 6. Chemical potential plot. The solid curve is from Eq.(27) and (28). The data sources (simulation results) are: Adam 1974[26]; Attard 1993 [27]; Heying&Corti 2004 [28]; Fixman 1983[29]; Heyes&Corti 2016 [30].

Figure 5. Isothermal compressibility of the hard sphere fluid. The MC simulation data are from Ref.[25].

## V Discussions and conclusions

The main assumption used in this work, Eq.(6), deserves some further discussions. The resultant EoS's, Eq.(13), (14) and (23) all diverge at $\eta = 1$. As a matter of fact, the same result can be obtained for any polynomial function of $k^{th}$ order. By assuming:

$$B_n = an^k + bn^{k-1} + \cdots + c \quad (30)$$

We have:

$$\lim_{n \to \infty} \frac{B_n}{B_{n+1}} = \frac{an^k + bn^{k-1} + \cdots + c}{a(n+1)^k + b(n+1)^{k-1} + \cdots + c} \to 1 \quad (31)$$

Yelash et al. (1999)[31] have proved that for the hard sphere fluid there exists a relation between the virial coefficients and a pole in the corresponding EoS, $\eta_{pole}$:

$$\eta_{pole} = \lim_{n \to \infty} \frac{B_n}{B_{n+1}} \quad (32)$$

Therefore, as long as the virial coefficients obey a polynomial function, the EoS built on it will always have an unphysical pole at $\eta_{pole} = 1$. In other words, if some physical pole (such as the close packing, $\eta_{pole} = 0.74048$) is expected, the polynomial relation, Eq.(30), must fail at some point. Taking the HS fluid as an example, Figure 3 shows that the quadratic relation, Eq.(24), holds up to $B_{16}$ and higher, should the extrapolations[19] be still valid. The breakdown of the Eq.(24) may happen if dramatic changes occur upon some virial coefficients. As conjectured by Ref[2], negative values may appear at some high order virial coefficients, such that a physically meaningful pole, $\eta_{pole} < 1$, can be expected.

Figures 2 and 4 show a common feature of the CS-type EoS's for both stable HD and HS fluids. Built on the virial coefficients available[2,5], the EoS's work well until the density approaches to the phase transition point. In the HD case, it is the liquid-hexact transition and in the HS case, the liquid-solid transition. For the HS fluid, several theories or suggestions have been proposed to address the phenomenon and the reader is referred to the reference[6,7,20,21] for details.

In conclusion, the new CS-type EoS's for the stable HD and HS fluids, Eq.(14) and Eq.(23) respectively, in particular the later, are simple and as accurate as the highest-level Padé approximations for compressibility predictions. For the applications where both simplicity and high accuracy are demanded, such as in the perturbation theory, they can be used as alternatives to, for instances, the CS EoS, Eq.(5), or the Kolafa EoS, Eq.(21).

## Data Availability

The data that support the findings of this study are available from the corresponding author upon reasonable request.